\documentclass[%
reprint,
superscriptaddress,
nofootinbib,
amsmath,
amssymb,
aps,
prd,
floatfix,
showkeys,
]{revtex4-2}

\usepackage{tabularx} 
\usepackage{multirow} 

\usepackage[normalem]{ulem}
\usepackage{fancyvrb}
\usepackage{fancyhdr}
\usepackage{graphicx,amsfonts,amssymb,amsbsy}
\usepackage{amsmath,amsthm,latexsym}
\usepackage[utf8]{inputenc}

\usepackage{natbib}
\usepackage[colorlinks]{hyperref}
\hypersetup{linkcolor=magenta,citecolor=blue}

\begin{document}

\title{Scaling and universality in strange quark stars}

\author{G. Lugones}
\email{german.lugones@ufabc.edu.br}
\affiliation{Universidade Federal do ABC, Centro de Ci\^encias Naturais e Humanas, Avenida dos Estados 5001- Bang\'u, CEP 09210-580, Santo Andr\'e, SP, Brazil.}

\author{A. G. Grunfeld}
\email{ag.grunfeld@conicet.gov.ar}
\affiliation{CONICET, Godoy Cruz 2290, Buenos Aires, Argentina} 
\affiliation{Departamento de F\'\i sica, Comisi\'on Nacional de Energ\'{\i}a At\'omica, Avenida del Libertador 8250, (1429) Buenos Aires, Argentina}

\begin{abstract}
We derive scaling laws that connect certain macroscopic observables of strange quark stars with key microscopic properties of self-bound quark matter, such as the energy per baryon at zero pressure and the strength of repulsive interactions. We also identify universal relations linking global properties of strange quark stars—specifically, their moment of inertia, tidal deformability, and both gravitational and baryonic compactness. Remarkably, these relations hold for two substantially different microscopic models—the quark-mass density-dependent model with excluded-volume corrections and the vector MIT bag model—underscoring their robust, model-independent nature. We demonstrate that the universal relations for strange quark stars differ significantly from those previously established for neutron stars composed of hadronic matter, thus enabling discrimination between the two types of objects without requiring detailed knowledge of their equations of state. Moreover, observational constraints on the maximum mass of compact stars could place bounds on both the depth of quark-matter self-binding and the strength of quark repulsive interactions.
\end{abstract}

\maketitle

\section{Introduction}  

Strange quark matter (SQM) is a hypothetical phase of deconfined quark matter composed of approximately equal numbers of up, down, and strange quarks. Its most intriguing property is its potential absolute stability, meaning it could be the true ground state of hadronic matter under certain conditions. This idea, originally proposed by Itoh \cite{Itoh:1970uw}, Bodmer \cite{Bodmer:1971we}, and Witten \cite{Witten:1984rs}, has driven extensive research into the possibility that strange quark matter could exist in nature, particularly in the form of compact astrophysical objects known as strange quark stars.

To describe the properties of SQM, various theoretical models have been developed. One of the most commonly employed frameworks is the MIT bag model. This approach enforces confinement through a bag constant, representing the extra energy needed to create a region of perturbative vacuum. Balancing the bag pressure against the degeneracy pressure of quarks enables the model to mimic both confinement and asymptotic freedom, providing a tractable description of quark matter in dense astrophysical environments~\cite{Farhi:1984qu}. Beyond MIT-based approaches, the Nambu--Jona-Lasinio (NJL) model offers a different perspective by employing a local four-fermion interaction to capture chiral symmetry breaking and restoration~\cite{Klevansky:1992qe}. Although the NJL model has successfully described various aspects of quark matter, its favored parameter sets generally do not predict stable strange quark matter; hence, it is not commonly used for strange quark stars~\cite{Buballa:1998pr}. An alternative description is provided by the quark mass density dependent (QMDD) model, where quark masses vary with the baryon number density \cite{Fowler:1981rp, Chakrabarty:1989bq, Chakrabarty:1991ui, Chakrabarty:1993db}. This variation effectively implements a dynamical confinement mechanism without requiring an explicit bag constant~ \cite{Benvenuto:1989kc, Lugones:1995vg, Benvenuto:1998tx, Lugones:2002vd, Peng:1999gh, Wang:2000dc, Peng:2000ff, Yin:2008me, Peng:2008ta, Xia:2014zaa}.

If strange quark matter is indeed stable, compact stars composed entirely of SQM—strange quark stars—should exist. These objects would differ significantly from neutron stars in several ways, including their mass-radius relations, cooling mechanisms, and gravitational wave signatures \cite{Alcock:1986hz, Page:2005fq, Weber:2004kj, Andersson:2001ev, Flores:2013yqa, VasquezFlores:2017uor}. 

The discovery of pulsars with masses around $2\,M_{\odot}$ \cite{Demorest:2010bx, Antoniadis:2013pzd, NANOGrav:2019jur, Riley:2021pdl, Miller:2021qha} placed significant constraints on the equation of state (EOS) of compact stars. Early versions of the MIT and QMDD models offered reasonable descriptions of SQM but generally predicted too soft an EOS to support such massive objects. To address this limitation, several modifications were introduced to stiffen the EOS, including color superconductivity \cite{Lugones:2002zd, Horvath:2004gn} and repulsive vector interactions \cite{Lopes2021_I,Lopes2021_II} in the MIT bag model. Similarly, variants of the QMDD model featuring density-dependent quark masses and additional repulsive effects were developed \cite{Xia:2014zaa, Li:2015ida, Issifu:2024zvq, You:2023bqx, Lugones:2023zfd, Lugones:2024ryz}. In particular, a QMDD model formulation incorporating an excluded volume correction effectively prevents quarks from occupying arbitrarily small volumes, thereby stiffening the EOS while preserving key features of the original model \cite{Lugones:2023zfd}. This mechanism substantially increases the maximum mass of strange quark stars, making the model consistent with observations of the most massive pulsars.

In this study, we carry out a comparative analysis of two distinct theoretical descriptions of SQM in compact stars: the QMDD model with excluded volume corrections and the vector MIT bag model (vMIT). Our main goals are:

\begin{itemize}  

\item Establish scaling relations that connect macroscopic observables—such as baryonic and gravitational mass, radius, moment of inertia, and tidal deformability—to microscopic quantities, including the energy per baryon of SQM at zero pressure and the strength of repulsive interactions, ensuring that these relations apply to both the QMDD and vMIT models.

Several studies have investigated similar relationships. For example, correlations have been found between macroscopic parameters (e.g., the maximum gravitational mass, maximum baryonic mass, and maximum moment of inertia) and fundamental properties such as the bag constant in the MIT bag model (see Ref.~\cite{Haensel:2007yy} and references therein). However, such scaling relations remain largely unexplored within the QMDD framework.

Moreover, to our knowledge, no existing work has sought to generalize these relations beyond specific model parameters by linking them to fundamental physical quantities independent of the chosen EOS. A particularly relevant example is the energy per baryon at zero pressure, which could serve as a more universal descriptor of strange quark matter properties, irrespective of the underlying model.

\item Explore universal relations between global properties of strange quarks stars that are independent of the chosen EOS. Universal relations in compact stars have been extensively investigated in recent years, revealing that certain combinations of macroscopic observables—such as the moment of inertia, tidal deformability, and quadrupole moment—can follow EOS-independent relationships \cite{Yagi:2013bca, Yagi:2013awa}. Numerous studies have extended these ideas, identifying further relations that involve non-radial oscillation frequencies and other stellar parameters \cite{Andersson:1997rn, VasquezFlores:2018tjl, Ranea-Sandoval:2022izm, Chirenti:2015dda}.

While our work is inspired by these previous findings, it differs in several important respects. First, whereas most earlier studies pursued universal relations applicable to hadronic or hybrid stars \cite{Kumar:2023ojk}, we focus exclusively on strange quark stars. This narrower scope enables us to derive relations that track numerical data with minimal dispersion. Second, unlike previous analyses of strange quark stars, which typically employed the standard MIT bag model without repulsive interactions, we adopt an extended version of the MIT bag model that includes vector repulsion, as well as the QMDD model—an entirely different theoretical framework. By systematically comparing these two models, we evaluate the robustness of universal relations under a broader range of assumptions for strange quark matter. Finally, we identify a previously unreported universal relation linking baryonic and gravitational mass, which was not present in earlier works. 
    
\end{itemize}

This paper is organized as follows. In Sec.~\ref{sec:EOS}, we detail the two quark-matter descriptions underlying our study. We begin by introducing the QMDD model, which incorporates an excluded volume correction to account for repulsive interactions, and then present the vector MIT bag model, emphasizing how vector repulsion stiffens the EOS. For both models, we identify the parameter space that renders quark matter self-bound and provide simple fits for the energy per baryon at zero pressure within this stability region. In Sec.~\ref{sec:stellar_structure}, we summarize the formalism for computing the mass-radius relation, tidal deformability, and moment of inertia under slow rotation, thereby characterizing the global properties of self-bound compact stars. Section~\ref{sec:scaling_relations} demonstrates that these macroscopic observables can be directly related to key microscopic EOS features, such as the energy per baryon at zero pressure and the repulsive interaction strength, through simple scaling laws. In Sect.~\ref{sec:universal_relations}, we show that these properties also obey universal relations, which include tidal deformability, moment of inertia, and both gravitational and baryonic compactness, which hold regardless of the chosen EOS. Some of these extend known results from the MIT bag framework, while others are new to the literature. Finally, in Sec.~\ref{sec:conclusions}, we summarize our findings and discuss their implications for the study of compact stars.

\section{Equations of state}
\label{sec:EOS}

In this section, we introduce the EOSs employed in this work.  Regardless of the chosen model, we treat the system as a mixture of quark flavors (\(u,d,s\)) and free electrons (\(e\)).  Our analysis focuses on the high–baryon–density, zero–temperature regime characteristic of cold compact stars, where first–principles QCD approaches face severe limitations: lattice QCD calculations are currently unable to explore this regime due to the sign problem, while perturbative QCD, are formally applicable only at asymptotically large \(\mu_B\gtrsim2.6\) GeV,  although they can nevertheless provide valuable constraints on the EOS at supranuclear densities (see e.g. \cite{Gorda:2023usm} and references therein).  Consequently, we rely on effective or phenomenological models to capture the essential physics of quark matter under these extreme conditions.

To ensure a self-consistent treatment, we impose local electric charge neutrality and chemical equilibrium under weak interactions. Since neutrinos freely escape the system ($\mu_{\nu_e} = 0$), the relevant chemical potentials satisfy
\begin{eqnarray}
\mu_d &=& \mu_u + \mu_e, \\
\mu_s &=& \mu_d,
\end{eqnarray}
and the charge neutrality condition reads
\begin{equation}
\tfrac{2}{3}\,n_u -\tfrac{1}{3}\,n_d -\tfrac{1}{3}\,n_s - n_e = 0.
\end{equation}
These relations fix the equilibrium composition of the system, which is examined in detail below.

In what follows,  we first introduce the QMDD model, highlighting its treatment of effective quark masses and potential non-perturbative corrections. We then describe the vMIT bag model and show how vector repulsion is incorporated. Next, we examine the parameter space for both models, identifying the stability windows where quark matter becomes self-bound. Finally, we provide simple parametric fits for constant energy-per-baryon curves within these stability regions.

\subsection{The quark-mass density-dependent model}

As discussed in Ref.~\cite{Lugones:2022upj}, the QMDD model treats quark matter as a quasiparticle system, where the energy density and the particle number densities retain the same functional forms as in the absence of interactions. Medium effects are incorporated through effective quark masses. Specifically, each quark mass consists of two parts: a flavor-dependent constant current mass $m_i$ and a flavor-blind term encoding medium contributions:
\begin{equation}
M_i = m_i + \frac{C}{n_B^{a/3}},
\label{eq:mass_ansatz}
\end{equation}
where $n_B$ is the baryon number density, $i$ indexes the quark flavors $u, d, s$, and $C$ and $a$ are free parameters. We take $m_{u} = m_d = 3.45~\mathrm{MeV}$, $m_{s} = 93~\mathrm{MeV}$, and treat electrons as massless. At low densities, $M_i$ diverges, effectively simulating confinement by preventing the presence of quasiparticles. Conversely, at extremely high densities, $M_i$ approaches the current quark mass, emulating chiral symmetry restoration in an effective manner when $m_i = 0$. The parameter $a$ is associated with the quark-quark interaction potential $v(r)\propto r^a$.

Because $M_i$ depends explicitly on $n_B$, the EOS is naturally formulated in the canonical ensemble~\cite{Lugones:2022upj}.
We note that a grand‐canonical formulation of the QMDD model is indeed feasible~\cite{Peng:2008ta,Xia:2014zaa}, but the explicit density dependence of the quark mass renders the thermodynamic relations more involved.  In the canonical ensemble, by contrast, $n_B$ is a natural variable and the density‐dependent mass can be treated directly as a function of the particle number and volume.  As shown in Ref.~\cite{Lugones:2024ryz}, both ensembles are formally equivalent, yet the canonical framework permits a straightforward implementation of excluded‐volume corrections. On the other hand, unlike other phenomenological models that embed thermodynamic variables in the Hamiltonian or Lagrangian (see e.g. Ref. \cite{Gorenstein:1995vm}), our formulation is entirely thermodynamic from the outset and does not require the retroactive introduction of corrective terms to restore consistency.  By starting from a well-defined Helmholtz free energy $F$, the standard derivatives of $F$ automatically guarantee full  thermodynamic consistency, and the confinement ``bag'' terms emerge naturally.

For point-like quasiparticles, the energy density $\epsilon_\mathrm{pl}$, the pressure $p_\mathrm{pl}$, and the chemical potentials $\mu_{\mathrm{pl},i}$ for each flavor follow Eqs.~\eqref{eq:pressure_general}, \eqref{eq:epsilon_general}, and \eqref{eq:mu_general}, respectively.

Owing to the dependence of the effective mass on $n_B$, additional terms arise in the pressure and chemical potentials relative to the free gas case. In particular, the extra term in the pressure is always negative and effectively acts like a bag constant, causing the pressure to vanish at a finite, large energy density. This feature allows the model to describe self-bound configurations.

Moreover, in effective models, one often introduces vector interactions that generate repulsive mean-field potentials; these are essential for reproducing the observed two-solar-mass pulsars. In the QMDD model, a similar stiffening of the EOS can be achieved by implementing an excluded-volume prescription, which reduces the available volume for quasiparticles and consequently increases the pressure at fixed density~\cite{Lugones:2023zfd}.

In practice, the excluded-volume correction is incorporated by replacing the system's volume $V$ with the available volume $\tilde{V}$, which excludes the volume $b(n_B)$ occupied by each quasibaryon:
\begin{equation}
\tilde{V} = V - b(n_B)\, n_B\, V.
\end{equation}
Defining the available volume fraction
\begin{equation}
q(n_B) \equiv \frac{\tilde{V}}{V} = 1 - n_B\, b(n_B),
\end{equation}
the mass formula, including excluded volume effects, becomes~\cite{Lugones:2023zfd}:
\begin{equation}
\tilde{M}_i = m_i + \frac{C}{\bigl(n_B / q\bigr)^{a/3}}.
\end{equation}
Assuming the ansatz
\begin{equation}
b = \frac{\kappa}{n_B},
\label{eq:definition_of_kappa}
\end{equation}
where $\kappa$ is a constant, it follows that
\begin{equation}
q = 1 - \kappa,
\end{equation}
and the thermodynamic quantities become
\begin{eqnarray}
\epsilon\bigl(\{n_j\}\bigr) & = & q \sum_i \epsilon_{\mathrm{pl}, i}\bigl(\{n_j / q\}\bigr) + e_e, \\
p\bigl(\{n_j\}\bigr) & = & q \sum_i p_{\mathrm{pl}, i}\bigl(\{n_j / q\}\bigr) + p_e, \\
\mu_k\bigl(\{n_j\}\bigr) & = & \mu_{\mathrm{pl}, k}\bigl(\{n_j / q\}\bigr).
\end{eqnarray}
The explicit forms of these expressions are given in Eqs.~\eqref{eq:epsilon_with_q}, \eqref{eq:p_with_q}, and \eqref{eq:mu_with_q}.

\begin{figure*}[tb]
\centering
\includegraphics[width=0.435 \linewidth]{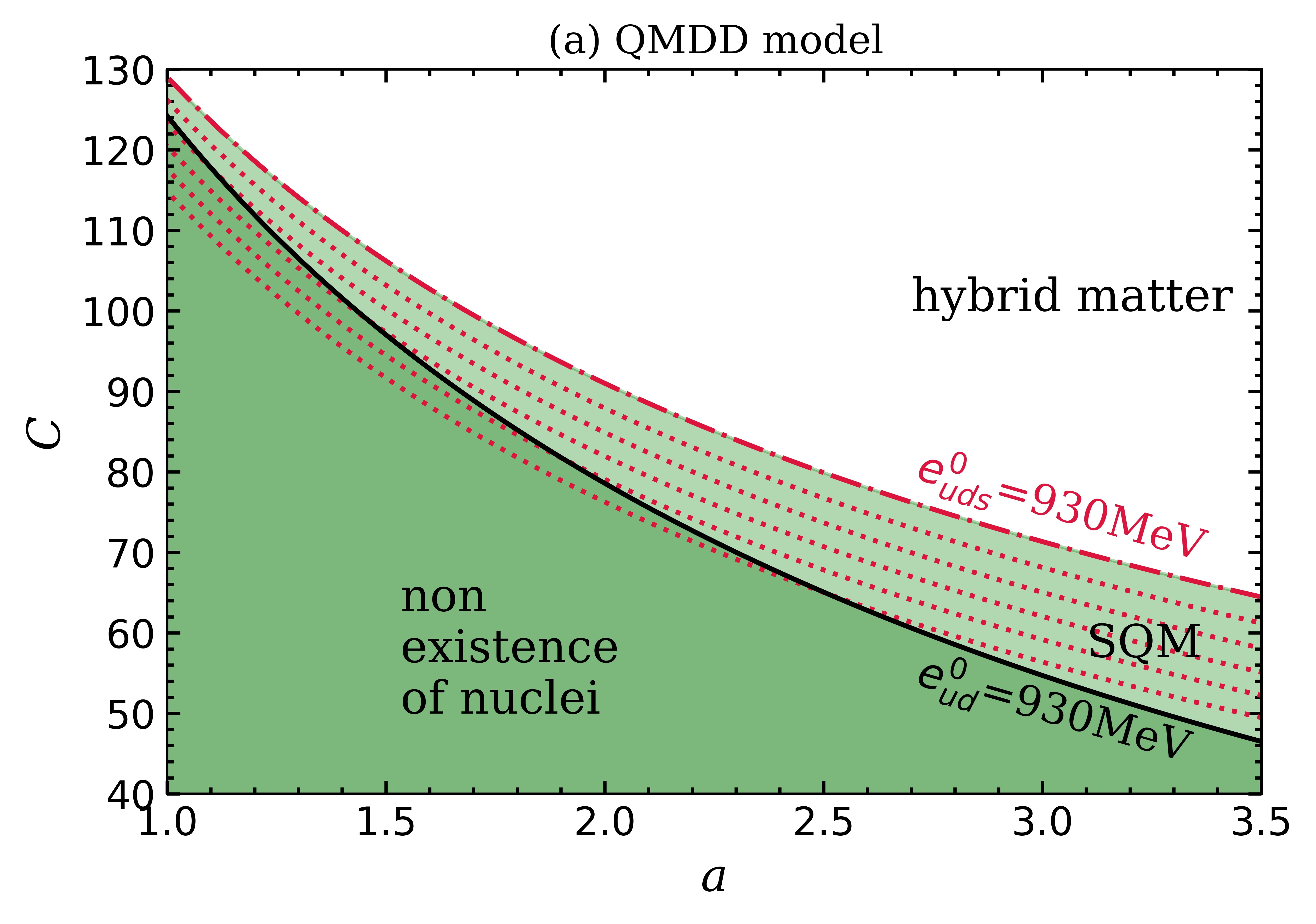}
\includegraphics[width=0.45 \linewidth]{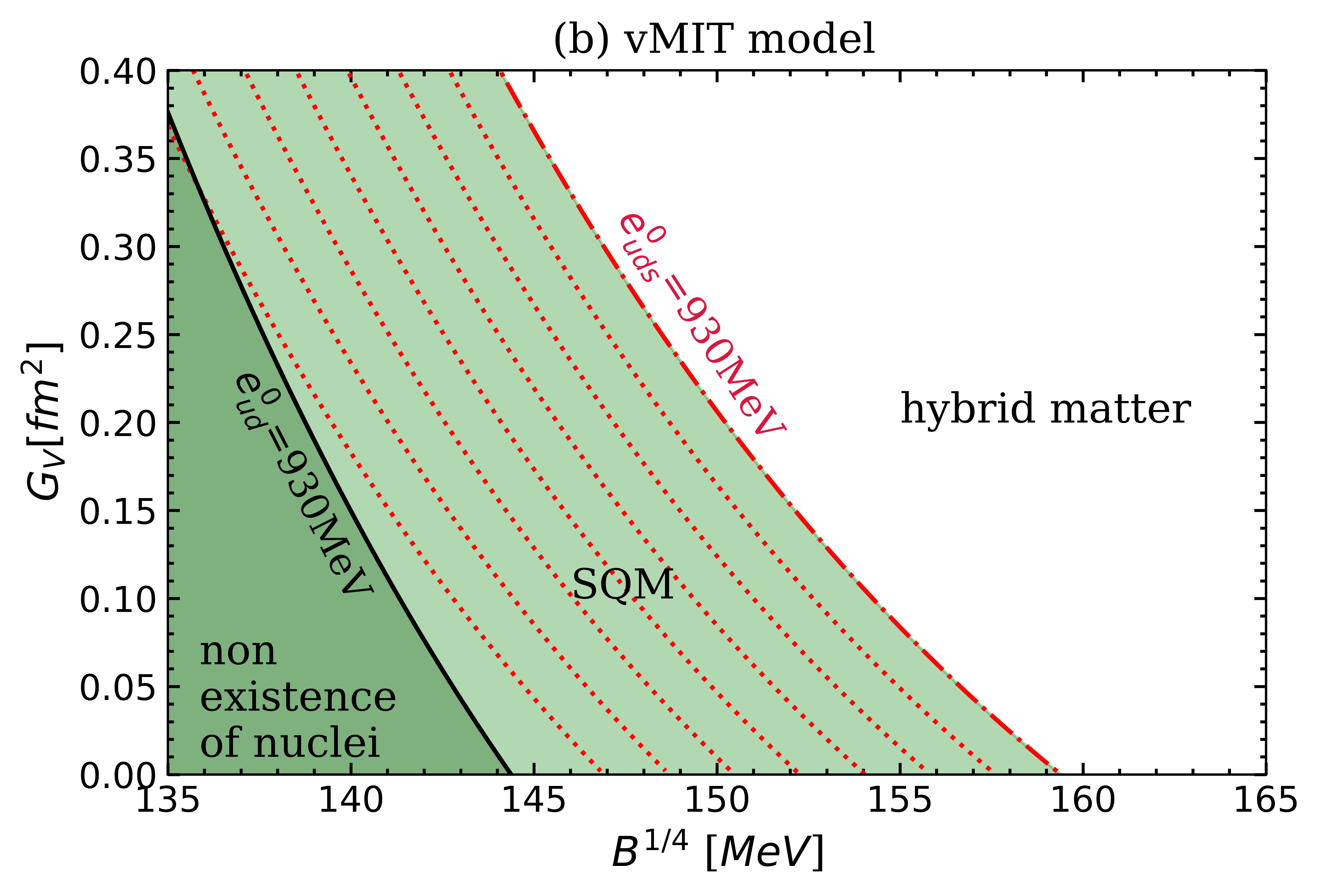}
\caption{Parameter space illustrating the bulk stability region of SQM for the QMDD model (a) and the vMIT model (b). The dotted red lines represent constant energy-per-baryon curves in 10\,MeV increments, ranging from 880\,MeV to 930\,MeV. The thick black line corresponds to $e^0_{ud} = 930$\,MeV. The light-green region denotes where SQM is energetically favored, whereas  the area labeled ``hybrid matter'' indicates parameter values that yield hybrid stars. The dark-green region represents parameter choices under which nuclei would become unstable.}
\label{fig:window}
\end{figure*}

\subsection{The vector MIT bag model}
\label{sec:quark_matter}

In this work, we adopt the model presented in Refs.~\cite{Lopes2021_I, Lopes2021_II}, characterized by the Lagrangian density
\begin{equation}
\begin{aligned}
\mathcal{L} & = \sum_{q}\Bigl[\bar{\psi}_{q}\bigl(i \gamma^{\mu} \partial_{\mu} - m_{q}\bigr) \psi_{q} - B\Bigr] \Theta(\bar{\psi}_{q} \psi_{q})  \\
& \quad + \sum_{q} g\,\bar{\psi}_{q}\gamma^{\mu} V_{\mu} \psi_{q}\,\Theta(\bar{\psi}_{q} \psi_{q}) 
+ \tfrac{1}{2} m_{V}^{2} V_{\mu} V^{\mu} \\
& \quad + \bar{\psi}_{e}\,\gamma_{\mu}\bigl(i \partial^{\mu}-m_{e}\bigr)\psi_{e},
\end{aligned}
\end{equation}
where $q$ stands for quarks ($u, d, s$), $e$ denotes electrons, and $B$ is the bag constant, representing the extra energy needed to create a region of perturbative vacuum per unit volume.  The Heaviside step function $\Theta$ equals one inside the bag and zero elsewhere. For simplicity, we assume a universal coupling $g$ between each quark and the vector field $V^{\mu}$, setting the vector-field mass to $m_{V} = 780\,\mathrm{MeV}$.

Within the mean-field approximation and defining $G_{V} \equiv \bigl(g/m_{V}\bigr)^{2}$, the vector-field equation of motion becomes
\begin{equation}
m_{V} V_{0}  = G_V^{1/2} \,(n_u + n_d + n_s),  
\end{equation}
where $n_{q}=\langle\bar{\psi}_{q}\,\gamma^{0}\,\psi_{q}\rangle$ is the quark number density. At zero temperature, the grand thermodynamic potential per unit volume is given by \cite{Lopes2021_II}
\begin{eqnarray}
\Omega & =& \sum_{q} \Omega^*_{q} + B - \tfrac{1}{2} m_{V}^{2} V_{0}^{2}  + \Omega_e,
\label{eq:Omega_bulk}
\end{eqnarray}
where $\Omega^*_{q}$ is the grand thermodynamic potential of a free Fermi gas of quarks: 
\begin{equation}
\Omega^*_{q} =  - g_q m_q^4 \phi(x_q),  \qquad g_q = 6,
\end{equation}
with
\begin{equation}
x_q =  \frac{\sqrt{\bigl(\mu_{q}^{*}\bigr)^{2} - m_q^{2}}}{m_q},
\end{equation}
being the dimensionless Fermi momentum, and the effective chemical potential defined as
\begin{eqnarray}
\mu_{q}^{*} =  \mu_{q} -  G_V^{1/2}\,m_V\,V_{0}.
\label{eq:effective_mu}
\end{eqnarray}
The function $\phi(x)$ is defined in Eq.~\eqref{eq_definition_of_phi}.  The grand thermodynamic potential of electrons is: 
\begin{equation}
\Omega_{e} =  - g_e m_e^4 \phi(x_e),  \qquad g_e = 2.
\end{equation}
From Eq.~\eqref{eq:Omega_bulk}, one can directly derive the EOS, and its explicit form can be found in Refs.~\cite{Lopes2021_I, Lopes2021_II}. In line with the QMDD model, we take $m_{u} = m_d = 3.45\,\mathrm{MeV}$, $m_{s} = 93\,\mathrm{MeV}$, and treat electrons as massless. The constants $B$ and $G_V$ are considered free parameters.

\subsection{Stability windows in the parameter space}

The energy per baryon of quark matter at zero pressure,
\begin{equation}
e^0 \equiv \frac{\epsilon}{n_B}\bigg|_{p=0},
\end{equation}
depends on the choice of EOS parameters and can lie above or below the binding energy per nucleon of the most stable atomic nucleus, $^{62}\mathrm{Ni}$, which is approximately 930~MeV.

If $e^0$ is below 930~MeV, quark matter is self-bound, meaning it does not convert in bulk to hadronic matter. In this scenario, compact stars would consist entirely of quark matter. Conversely, when $e^0$ exceeds 930~MeV, the matter is considered hybrid. In this case, there is a transition from hadronic matter at low pressures to deconfined quark matter at higher pressures, yielding hybrid stars with a quark matter core surrounded by hadronic layers.

Figures~\ref{fig:window}(a) and \ref{fig:window}(b) illustrate the stability regions for quark matter in the QMDD and vMIT models. In the QMDD model, three parameters govern the behavior: $a$, $C$, and $b$. However, as shown in Ref.~\cite{Lugones:2023zfd}, the stability region is unaffected by $b$, which specifies the excluded volume per baryon. Thus, the stability window depends only on $a$ and $C$. In the vMIT model, the stability region is determined by the parameters $B$ and $G_V$, since quark masses are held fixed.
Each plot highlights two key curves. The solid black line, labeled $e_{ud}^0 = 930~\mathrm{MeV}$, corresponds to the parameter combinations for which the energy per baryon in two-flavor quark matter equals 930~MeV. The dashed red line, labeled $e_{uds}^0 = 930~\mathrm{MeV}$, indicates the parameter values where the energy per baryon in three-flavor quark matter is exactly 930~MeV. The light-green shaded region between these curves denotes parameter values for which three-flavor quark matter is more stable than the most tightly bound atomic nucleus, while two-flavor quark matter remains less stable—a scenario consistent with the strange quark matter hypothesis. EOS models defined by these parameters permit self-bound quark matter without contradicting nuclear stability. Within this region, the dotted red curves represent $e_{uds}^0$ values ranging from 930~MeV down to 880~MeV, in increments of 10~MeV. The dotted red curves can be described by simple fits in terms of the EOS parameters, as presented in Appendix~\ref{sec:appendix_B}.

The dark-green region corresponds to parameter values for which two-flavor quark matter in bulk is more stable than the most tightly bound nucleus, implying that nuclei could spontaneously decay into a single deconfined bag of $u$ and $d$ quarks—an outcome in clear conflict  with observational evidence. For both the QMDD and vMIT models, the energy per baryon in three-flavor matter remains consistently lower than in two-flavor matter throughout the stability window; hence, these models do not support the absolute stability of two-flavor quark matter.

\section{Stellar structure}
\label{sec:stellar_structure} 

The microscopic models introduced above will be used to determine several global properties of self-bound stars, such as the mass-radius relation, tidal deformability, and moment of inertia under slow rotation conditions. For completeness, we summarize in this section the equations required to compute these properties. In the next section, we will derive universality and scaling relations involving these quantities.

\subsection{TOV equations}

We begin with the metric of a stationary, spherically symmetric spacetime,
\begin{equation}
ds^2= dt^2 e^{2 \nu}-e^{2 \lambda} d r^2-r^2 (d \theta^2+\sin ^2 \theta d \phi^2),
\end{equation}
where $t$ is the time coordinate,  $r$ is a radial coordinate, $\theta$ and $\phi$ are the polar and azimuthal angles, respectively, while $\nu$ and $\lambda$ are some functions of $r$. In the perfect-fluid approximation, the hydrostatic equilibrium of such configurations is described by the Tolman-Oppenheimer-Volkoff (TOV) equations \cite{Tolman1939, OppenheimerVolkoff1939}, which can be solved once an EOS of the form $p = p(\epsilon)$ is specified:
\begin{eqnarray}
\frac{d p}{d r} & = &-\frac{\epsilon m}{r^2}\left(1+\frac{p}{\epsilon }\right)\left(1+\frac{4 \pi p r^3}{m }\right)\left(1-\frac{2 m}{ r}\right)^{-1}, \quad \label{eq:TOV_p}\\
\frac{d m}{d r} & = & 4 \pi r^2 \epsilon ,\\
\frac{d \nu}{d r} & =& -\frac{1}{\epsilon } \frac{d p}{d r}\left(1+\frac{p}{\epsilon }\right)^{-1} ,
\end{eqnarray}
where the gravitational mass $m$ inside radius $r$ is defined by
\begin{equation}
e^{-\lambda} \equiv \sqrt{1-2 m /r }.
\end{equation}
For further details, see Ref. \cite{Haensel:2007yy}. The stellar surface is defined by the condition $p(R)=0$, being $R$ the stellar radius. The gravitational mass of the star is then
\begin{equation}
M_G \equiv m(R) . 
\end{equation}

The baryon number for a canonical neutron star $\left(M_G=1.4 M_{\odot}\right)$ can be estimated as $A_B \simeq 1.4 M_{\odot} / m_n=1.7 \times 10^{57}$. It is useful to define the baryon mass of the star, denoted as $M_B \equiv A_B m_B$, where $m_B$ represents the mass of one baryon. A common choice for $m_B$ is to set it equal to the neutron mass, $m_n$. This assumption leads to the equation
\begin{equation}
M_B=A_B m_n \simeq 0.842 A_{\mathrm{B} 57} M_{\odot}
\end{equation}
where $A_{\mathrm{B} 57} \equiv A_B / 10^{57}$. The values of $A_B$ and $M_B$ remain constant throughout the evolutionary path of an isolated star.

\subsection{Dimensionless tidal deformability}

During the inspiral phase of a neutron star-neutron star merger, tidal forces cause the stars to deform significantly, producing an observable effect on the gravitational wave signal emitted by the system. This deformation is quantified by the dimensionless tidal deformability, a parameter that measures how the star responds to an external tidal field.  The relationship between the induced mass quadrupole moment  $Q_{ij}$ and the external tidal field $\epsilon_{ij}$ is given by $Q_{ij} = -\Lambda M_G^5 \epsilon_{ij}$, where $\Lambda$ denotes the tidal deformability. 

The tidal deformability $\Lambda$ can be computed via
\begin{equation}
\Lambda=\frac{2}{3} k_2 \mathcal{C}_G^{-5},
\label{eq:definition_of_Lambda}
\end{equation}
where $\mathcal{C}_G = M_G / R$ is the star's compactness.
The parameter $k_2$ is known as the second-order Love number and is given by 
\begin{equation}
\begin{aligned} 
k_{2}= & \frac{8 C^{5}}{5}(1-2 C)^{2}\left[2-y_{R}+2 C\left(y_{R}-1\right)\right] \\ & \times\left\{2 C\left[6-3 y_{R}+3 C\left(5 y_{R}-8\right)\right]\right. \\ & +4 C^{3}\left[13-11 y_{R}+C\left(3 y_{R}-2\right)+2 C^{2}\left(1+y_{R}\right)\right] \\ & \left.+3(1-2 C)^{2}\left[2-y_{R}+2 C\left(y_{R}-1\right)\right] \ln (1-2 C)\right\}^{-1}
\end{aligned}
\end{equation}
where $y_R=y(r=R)$ and $y(r)$ is obtained by solving:
\begin{equation}
 r\frac{dy}{dr} +y^2 + yF(r) +r^2Q(r) = 0 . \label{EL15}
\end{equation}
The coefficients $F(r)$ and $Q(r)$ are given by:
\begin{eqnarray} 
F(r) &= & {\left[1-4 \pi r^{2}(\epsilon-p)\right]\left[1-\frac{2 m}{r}\right]^{-1} } , \\ 
Q(r) & = & 4 \pi\left[5 \epsilon+9 p+\frac{\epsilon+p}{c_{s}^{2}}-\frac{6}{4 \pi r^{2}}\right]\left[1-\frac{2 m}{r}\right]^{-1}   \nonumber \\ 
& & -\frac{4 m^{2}}{r^{4}}\left[1+\frac{4 \pi r^{3} p}{m}\right]^{2}\left[1-\frac{2 m}{r}\right]^{-2} ,
\label{EL17}
\end{eqnarray}
where $c_{s}^{2} \equiv d p / d \epsilon$ is the squared speed of sound. The boundary condition for Eq. \eqref{EL15}  at $r = 0$ is given by $y(0) = 2$. To obtain the tidal Love number, we use the EOS of Sec. \ref{sec:EOS} and integrate the TOV equations along with Eq. \eqref{EL15}. For further details, see Ref.~\cite{Postnikov:2010yn}.

\subsection{Moment of inertia for slow rigid rotation}

The effects of rotation on the structure of compact stars are commonly treated within the slow-rotation (perturbative) approximation~\cite{Hartle:1967he}. In this work, we restrict ourselves to slow, uniform (rigid) rotation with an angular frequency $\Omega$, as measured by a distant observer.

Rotation induces polar flattening, leading to an axially symmetric configuration. To leading order in $\Omega$, the total angular momentum $J$ is proportional to $\Omega$, so that the moment of inertia,
\begin{equation}
I = \frac{J}{\Omega},
\end{equation}
depends solely on the mass distribution and spacetime geometry of the corresponding nonrotating star~\cite{Hartle:1967he}. In this approximation, the moment of inertia $\mathcal{I}(r)$ of the matter enclosed within a radius $r$ is obtained by solving the differential equation~\cite{Hu:2023vsq, Dong:2023vxv}
\begin{equation}
\frac{d\mathcal{I}}{dr} = \frac{8}{3}\pi r^4 \epsilon \left(1+\frac{p}{\epsilon}\right) \left(1-\frac{5}{2}\frac{\mathcal{I}}{r^3}+\frac{\mathcal{I}^2}{r^6}\right) \left(1-\frac{2m}{r}\right)^{-1}
\label{eq:moment_of_inertia}
\end{equation}
and the star's total moment of inertia is
\begin{equation}
I \equiv \mathcal{I}(R),
\end{equation}
where $R$ is the stellar radius.

\section{Scaling relations for global strange quark star properties}
\label{sec:scaling_relations}

In this section, we show that key macroscopic properties of strange quark stars—such as mass, radius, moment of inertia, and tidal deformability—can be directly linked to fundamental microscopic parameters. We focus on two main aspects:
\begin{itemize}
    \item The maximum mass of a stellar sequence is found to be directly related to the energy per baryon of quark matter at zero pressure, $e^0$.
    
    \item Several global properties of strange quark stars depend in a straightforward manner on parameters that quantify repulsive interactions. In the QMDD model, these interactions are encoded in the excluded volume fraction $q$, whereas in the vMIT model they are characterized by the coupling constant $G_V$. Unlike the scaling with $e^0$, the fits involving these repulsive parameters remain valid for stars of any mass or central pressure.
\end{itemize}

Some of these relationships can be derived analytically, while others are obtained by fitting numerical results.

\subsection{Scaling properties in the QMDD model}

\subsubsection{Excluded volume}

To derive a scaling relation involving the excluded volume fraction $q$, we first note that $q$ does not explicitly appear in the electron contributions to the energy density and pressure, as seen in Eqs.~\eqref{eq:epsilon_with_q} and \eqref{eq:p_with_q}. To factor out the quantity $q$ in the EOS, we approximate these expressions by neglecting the electron contribution (while still including it when enforcing charge neutrality and chemical equilibrium), so that
\begin{equation}
\epsilon = q\,\epsilon_\mathrm{pl} \quad \text{and} \quad p = q\,p_\mathrm{pl},
\end{equation}
where $\epsilon_\mathrm{pl}$ and $p_\mathrm{pl}$ are the energy density and pressure for point-like particles.

Substituting these modified expressions into Eq.~\eqref{eq:TOV_p} and assuming that $q$ is constant, we obtain
\begin{eqnarray}
\frac{d p_\mathrm{pl}}{d r} & = & -\frac{\epsilon_\mathrm{pl} m}{r^2} \left(1+\frac{p_\mathrm{pl}}{\epsilon_\mathrm{pl}}\right) \left(1+\frac{4 \pi q\,p_\mathrm{pl}\,r^3}{m}\right) \nonumber \\
&& \times \left(1-\frac{2m}{r}\right)^{-1}.
\label{eq:TOV_pl}
\end{eqnarray}
Next, we define the effective variables $r_\mathrm{pl}$ and $m_\mathrm{pl}$ such that
\begin{equation}
q\,\frac{r^3}{m} = \frac{r_\mathrm{pl}^3}{m_\mathrm{pl}}, \qquad \frac{m}{r} = \frac{m_\mathrm{pl}}{r_\mathrm{pl}}.
\label{eq:scaling_q_1}
\end{equation}
Substituting these definitions into Eq.~\eqref{eq:TOV_pl} shows that the TOV equation retains the same form as in the point-like case, with $r_\mathrm{pl}$ and $m_\mathrm{pl}$ representing the radial coordinate and enclosed mass, respectively, when the structure equations are integrated using an EOS that does not account for the excluded volume.

From Eq.~\eqref{eq:scaling_q_1}, the following scaling relations immediately follow:
\begin{equation}
r = \frac{r_\mathrm{pl}}{\sqrt{q}}, \qquad m = \frac{m_\mathrm{pl}}{\sqrt{q}}.
\label{eq:scaling_q_2}
\end{equation}
This scaling relation can be interpreted as follows. For any given central pressure, the mass and radius of a star can be calculated once using the point-like EOS, without corrections due to repulsive interactions. To obtain the mass–radius relation when excluded-volume effects are included, it is not necessary to re-solve the TOV equations; one simply applies the correction factor $\sqrt{q}$ to the results.

Using a similar procedure for Eq. \eqref{eq:moment_of_inertia}, it is straightforward to show that:
\begin{equation}
\mathcal{I} =  \frac{\mathcal{I}_\mathrm{pl}}{q^{3/2}}   
\label{eq:scaling_I}
\end{equation}
where $\mathcal{I}$ and $\mathcal{I}_\mathrm{pl}$ are the moments of inertia with and without considering the excluded volume, respectively.

Finally, let us analyze the scaling of the tidal deformability. It can be easily verified that both $k_2$ and $\mathcal{C}_G$ remain invariant under changes in $q$. Therefore, using Eq.~\eqref{eq:definition_of_Lambda}, we find that $\Lambda$ is also invariant with respect to changes in $q$:
\begin{equation}
\Lambda =  \Lambda_\mathrm{pl}.
\label{eq:scaling_II}
\end{equation}

Notably, even though equations \eqref{eq:scaling_q_2}, \eqref{eq:scaling_I}, and \eqref{eq:scaling_II} were derived analytically by neglecting the contribution of electrons, they still accurately reproduce the results obtained using the complete EOS.

\begin{figure}[tb]
\centering
\includegraphics[width=\linewidth]{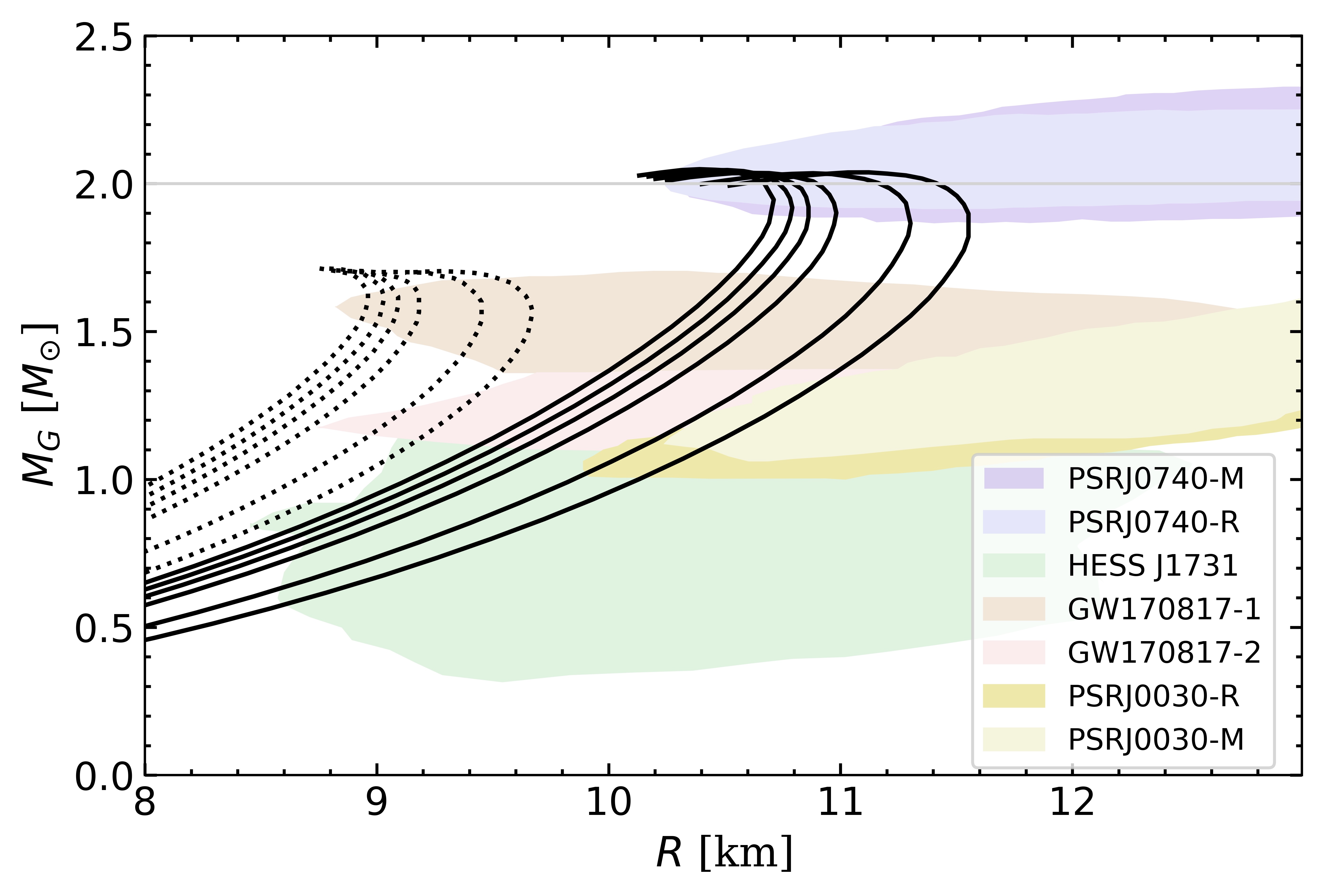}
\caption{Mass--radius relations for strange quark stars in the QMDD model computed by selecting $a$ and $C$ along the $e_{uds}^0 = 900$\,MeV curve from Fig.~\ref{fig:window}, ensuring these parameters lie within the stability region (light green area). Solid lines correspond to $\kappa = 0.3$ ($q=0.7$), while dotted lines represent $\kappa = 0$ ($q=1$). Each set of curves yields the same maximum mass for fixed $e_{uds}^0$ and $q$. Colored bands indicate observational constraints from various neutron-star measurements. }
\label{fig:MR_1}
\end{figure}

\begin{table}[tbh]
\centering
\renewcommand{\arraystretch}{1.3} 
\begin{tabularx}{0.9\columnwidth}{|c|c|X|X|X|X|X|X|}
\hline
\multicolumn{2}{|c|}{\multirow{2}{*}{}} & \multicolumn{6}{c|}{$e_{uds}^0$ [MeV]} \\ \cline{3-8} 
\multicolumn{2}{|c|}{}                  & 930    & 920    & 910    & 900    & 890    & 880    \\ \hline
\multirow{7}{*}{\rotatebox{90}{Volume fraction $q$}}  
& 1.0 & 1.596 & 1.633 & 1.669 & 1.707 & 1.746 & 1.787 \\
& 0.9 & 1.684 & 1.721 & 1.760 & 1.799 & 1.841 & 1.882 \\ 
& 0.8 & 1.786 & 1.826 & 1.866 & 1.908 & 1.950 & 1.996 \\ 
& 0.7 & 1.909 & 1.949 & 1.994 & 2.037 & 2.086 & 2.133 \\ 
& 0.6 & 2.060 & 2.105 & 2.153 & 2.201 & 2.252 & 2.305 \\ 
& 0.5 & 2.256 & 2.306 & 2.358 & 2.410 & 2.467 & 2.524 \\ 
& 0.4 & 2.523 & 2.580 & 2.635 & 2.697 & 2.759 & 2.822 \\ \hline
\end{tabularx}
\caption{Maximum mass (in units of $M_{\odot}$) for strange quark stars in the QMDD model, tabulated as a function of the energy per baryon at zero pressure, $e_{uds}^0$ (columns), and the volume fraction $q$ (rows). }
\label{table:compact_star_mass}
\end{table}

\begin{figure}[h]
\centering
\includegraphics[width=0.8\linewidth]{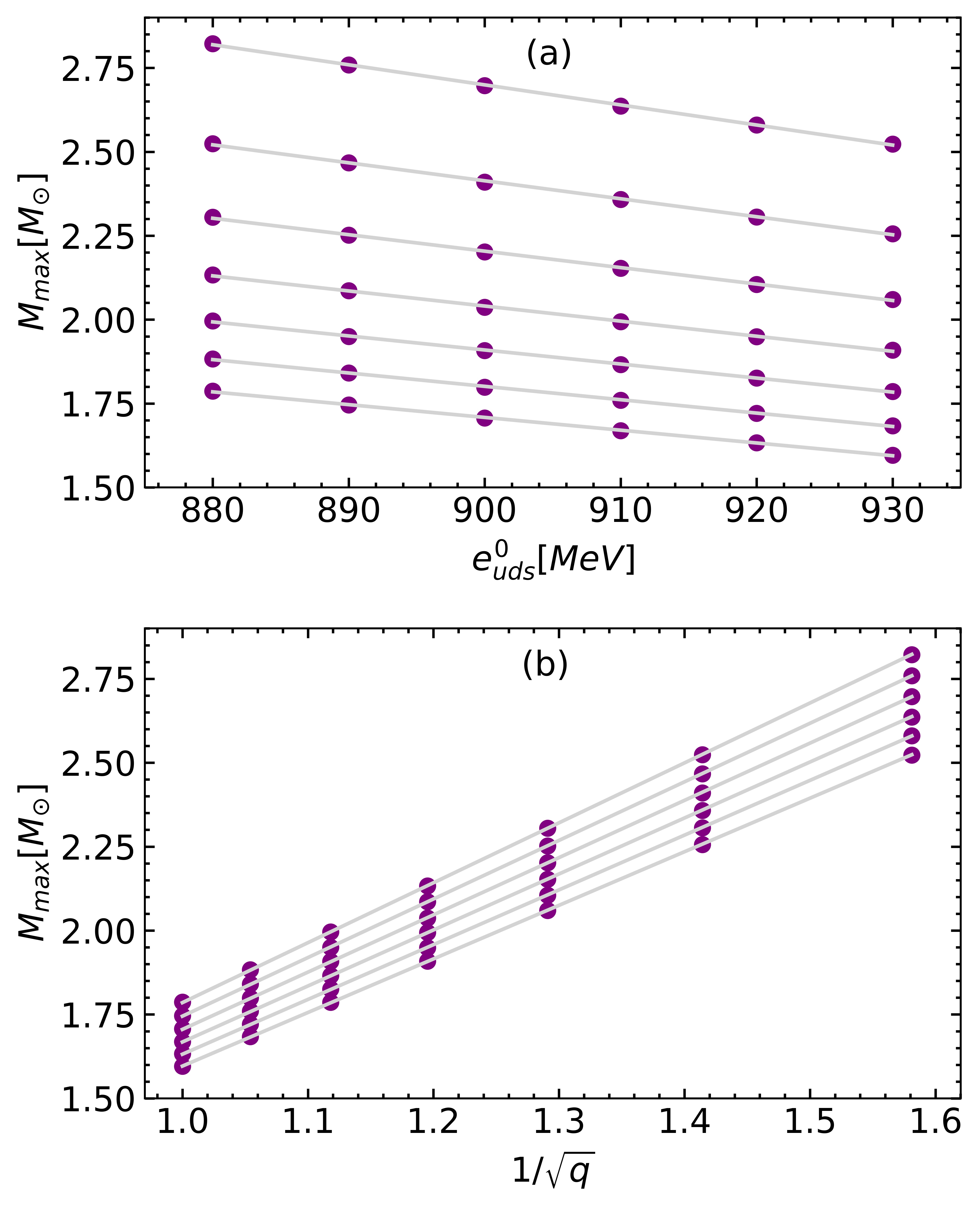}
\caption{(a) Linear fits of the maximum mass in the QMDD model as a function of the energy per baryon at zero pressure, $e_{uds}^0$, for different volume fractions $q$. The data are taken from Table~\ref{table:compact_star_mass}, showing that $M_{\mathrm{max}}$ increases linearly with $e_{uds}^0$ for a fixed $q$, ranging from $q=0.4$ (upper curve) to $q=1.0$ (lower curve). (b) Linear fits of $M_{\mathrm{max}}$ as a function of $1/\sqrt{q}$ for various $e_{uds}^0$ values, spanning 930\,MeV (lowest curve) to 880\,MeV (highest curve). }

\label{fig:Mmax_fits}
\end{figure}

\subsubsection{Energy per baryon}

Figure~\ref{fig:MR_1} illustrates the mass--radius relation for strange quark stars computed by selecting $a$ and $C$ along a constant-$e_{uds}^0$ curve, ensuring that the parameters lie within the SQM region. As an example, we set $e_{uds}^0 = 900\,\mathrm{MeV}$ and consider two values of $\kappa$: 0.3 ($q=0.7$) and 0 ($q=1$). In total, we show 12 curves, corresponding to 6 distinct combinations of $a$ and $C$ (all sharing the same $e_{uds}^0$) and the two $\kappa$ values. Although the shapes of these curves differ within each $\kappa$ family, they yield the same maximum mass, indicating that the maximum mass exhibits a scaling relation with both $e_{uds}^0$ and $q$.

Table~\ref{table:compact_star_mass} lists the maximum masses for various parameter choices. Using these data, Fig.~\ref{fig:Mmax_fits}(a) demonstrates a linear dependence of $M_{\mathrm{max}}$ on $e_{uds}^0$ for fixed $q$, whereas Fig.~\ref{fig:Mmax_fits}(b) shows a linear dependence on $1/\sqrt{q}$ for fixed $e_{uds}^0$, consistent with the prediction of Eq.~\eqref{eq:scaling_q_2}. The results can be summarized by the following fit:
\begin{equation}
M_{\mathrm{max}} = \frac{1}{\sqrt{q}} \left[5.1098  -3.5154 \left( \frac{e^0_{uds}}{930 \, \mathrm{MeV}}  \right) \right] M_{\odot}  ,
\label{eq:fit_MMAX_QMDD}
\end{equation}
which reproduces the values in Table~\ref{table:compact_star_mass} with an error smaller than 0.2\%.

\subsection{Scaling properties in the  vMIT}

\begin{figure}[tb]
\centering
\includegraphics[width=0.8\linewidth]{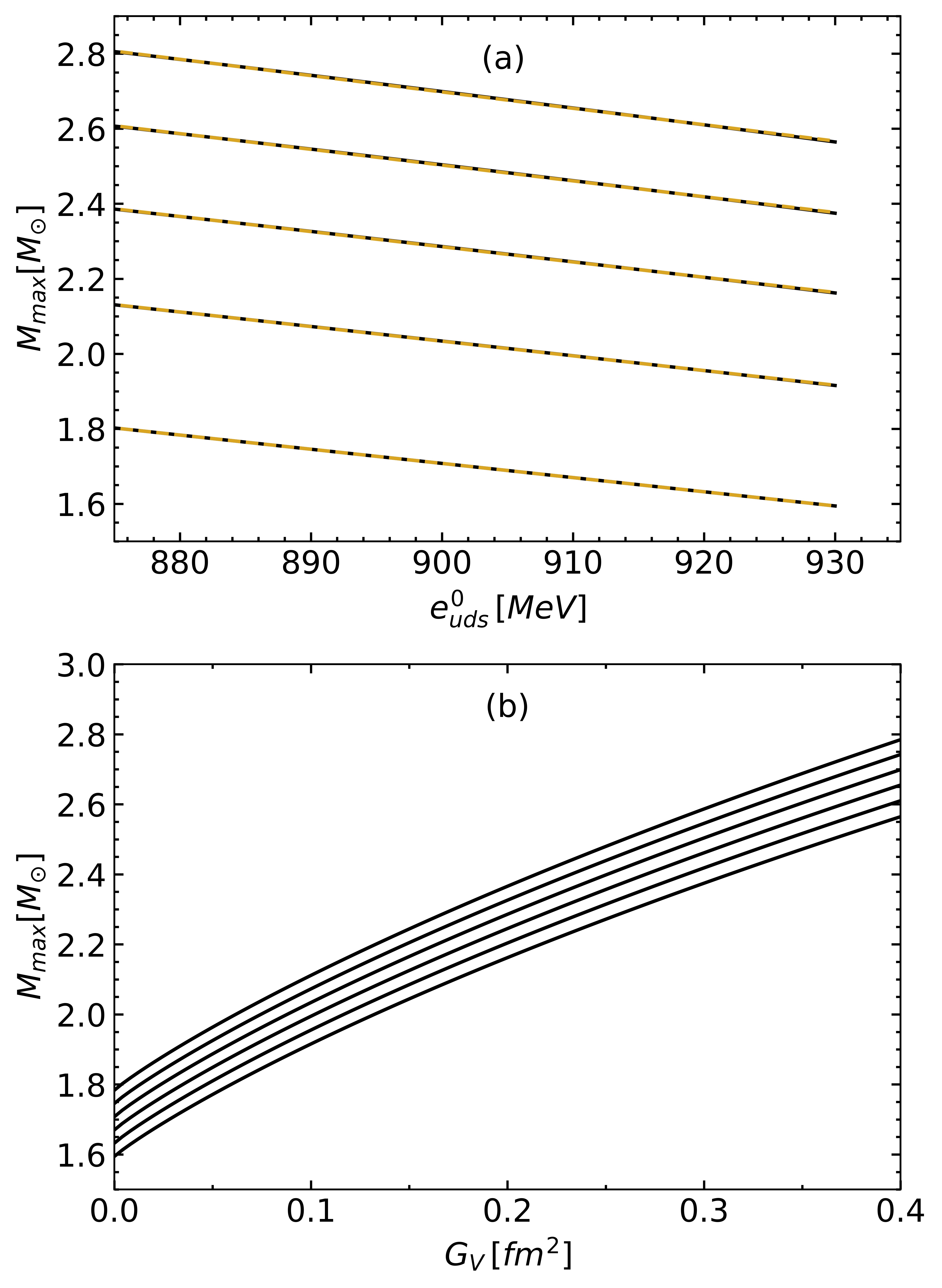}
\caption{(a) Maximum mass of strange quark stars in the vMIT model as a function of $e^0_{uds}$ for five fixed values of $G_V$ (0.0, 0.1, 0.2, 0.3, 0.4), with linear fits shown as dashed lines. (b) Maximum mass versus $G_V$ for $e^0_{uds}$ ranging from 880\,MeV to 930\,MeV in increments of 10\,MeV. These results mirror the trends observed in Fig.~\ref{fig:Mmax_fits}, demonstrating that $M_{\mathrm{max}}$ scales linearly with $e^0_{uds}$ and also varies systematically with the strength of repulsive interactions, here quantified by $G_V$.}
\label{fig:Mmax_fits_4ab}
\end{figure}

\begin{table}[tbh]
\centering
\renewcommand{\arraystretch}{1.3} 
\begin{tabularx}{0.9\columnwidth}{|c|c|X|X|X|X|X|}
\hline
\multicolumn{2}{|c|}{\multirow{2}{*}{}} & \multicolumn{5}{c|}{$e_{uds}^0$ [MeV]} \\ \cline{3-7} 
\multicolumn{2}{|c|}{}                  & 930    & 920    & 910    & 900    & 890    \\ \hline
\multirow{5}{*}{\rotatebox{90}{$G_V \, [\mathrm{fm}^2]$}}  
& 0.0 & 1.611 & 1.646 & 1.683 & 1.721 & 1.760 \\ 
& 0.1 & 1.938 & 1.974 & 2.012 & 2.051 & 2.091 \\ 
& 0.2 & 2.192 & 2.231 & 2.269 & 2.309 & 2.351 \\ 
& 0.3 & 2.407 & 2.445 & 2.485 & 2.530 & 2.574 \\ 
& 0.4 & 2.595 & 2.637 & 2.680 & 2.724 & 2.769 \\ \hline
\end{tabularx}
\caption{Maximum mass $M_{\mathrm{max}}$ (in units of $M_{\odot}$) for strange quark stars in the vMIT model, listed as a function of the energy per baryon at zero pressure, $e_{uds}^0$, and the coupling constant $G_V$.}
\label{table:compact_star_mass_2}
\end{table}

We now apply the same scaling analysis introduced for the QMDD model to the vMIT framework, where repulsive interactions are governed by the coupling constant $G_V$ rather than an excluded volume fraction $q$. Figures~\ref{fig:Mmax_fits_4ab}(a) and \ref{fig:Mmax_fits_4ab}(b) show how the maximum mass depends on both $e^0_{uds}$ and $G_V$. Table~\ref{table:compact_star_mass_2} lists $M_{\mathrm{max}}$ for various combinations of $e^0_{uds}$ and $G_V$.

A key similarity to the QMDD model is evident when $G_V = 0$: the vMIT results coincide with those of the QMDD model in the no-excluded-volume limit ($q=1$). As $G_V$ increases, both the slope and the intercept of the $M_{\mathrm{max}}$ versus $e^0_{uds}$ lines shift, reflecting the growing contribution of repulsive interactions. In particular, for each fixed $G_V$, $M_{\mathrm{max}}$ lies along a straight line as a function of $e^0_{uds}$, mirroring the linear relationship observed in the QMDD model for fixed $q$.

These findings can be summarized by the analytic fit
\begin{equation}
M_{\mathrm{max}} = \frac{1}{\sqrt{q^\star}} \left[5.1098  -3.5154 \left( \frac{e^0_{uds}}{930 \mathrm{MeV}}  \right) \right] M_{\odot}  ,
\label{eq:fit_MMAX_vMIT}
\end{equation}
where
\begin{equation}
\frac{1}{\sqrt{q^\star}} \equiv  1  +  1.3\,G_V^{0.8}.
\end{equation}
This expression reproduces the data in Table~\ref{table:compact_star_mass_2} with better than 2\% accuracy. Although scaling relations for linear EOSs have been discussed previously~\cite{Haensel:2007yy}, the above result applies to the more complex EOS of the vMIT model, underscoring the parallel between the role of $G_V$ here and that of $q$ in the QMDD model.

\subsection{Comparing maximum-mass fits in the QMDD and vMIT models}

A noteworthy outcome of the previous analysis is the striking similarity between the final scaling relations derived in the QMDD and vMIT models. In the QMDD model, the maximum mass is given by Eq.~\eqref{eq:fit_MMAX_QMDD}, whereas in the vMIT model it is given by Eq.~\eqref{eq:fit_MMAX_vMIT}. Remarkably, the factor
\begin{equation}
\left[5.1098  -3.5154 \left( \frac{e^0_{uds}}{930 \mathrm{MeV}}  \right) \right] M_{\odot}
\end{equation}
is identical in both models, despite their distinct underlying assumptions. 

Another notable similarity is that, in both models, the repulsive interactions appear as a global multiplicative factor: $\tfrac{1}{\sqrt{q}}$ in the QMDD model and $\tfrac{1}{\sqrt{q^\star}}$ in the vMIT model. In the former, $q$ directly represents an excluded volume fraction, whereas $q^\star$ in the latter does not strictly measure an excluded volume but still encodes the net repulsive effect through $G_V$. Physically, both $q$ and $q^\star$ stiffen the EOS thereby raising the maximum mass. Even though $q^\star$ is not literally a volume fraction, it captures the same fundamental tendency to keep quarks from packing too closely, mirroring the role of $q$ in the QMDD model.

Hence, while the QMDD and vMIT models implement repulsive interactions in distinct ways, they both exhibit a similar scaling structure for $M_{\mathrm{max}}$, highlighting a common physical mechanism by which repulsive forces increase the maximum mass of strange quark stars. Unfortunately, scaling relations analogous to those in Eqs.~\eqref{eq:scaling_q_2}, \eqref{eq:scaling_I}, and \eqref{eq:scaling_II} do not emerge in the vMIT model.

\section{Universal relations}
\label{sec:universal_relations}

Universal relations link macroscopic observables of compact stars—such as tidal deformability, moment of inertia, or compactness—independently of the specific EOS. Such relations have been extensively documented for hadronic stars \cite{Yagi:2016bkt}, whereas comparatively fewer studies exist for quark-containing stars, most of which rely on simplified versions of the MIT bag model \cite{Yagi:2016bkt}. In this work, we demonstrate that some of these relations extend to significantly different descriptions of strange quark matter, including the QMDD model. Moreover, we identify additional universal relationships specific to strange quark stars that have not been reported previously.

\subsection{Moment of inertia versus compactness}
\label{sec:I_vs_C}

A number of studies have explored how the dimensionless moment of inertia $I/(M_G R^2)$ for slowly rotating stars can be related to the stellar compactness  via low-order polynomial expansions with only weak EOS dependence \cite{Ravenhall1994ApJ, Lattimer:2000nx, Bejger:2002ty, Lattimer:2004nj, Greif:2020pju}. For instance, \cite{Lattimer:2004nj} proposed
\begin{equation}
\frac{I}{M_G R^2} = \tilde{a}_0 + \tilde{a}_1 \mathcal{C}_G + \tilde{a}_4 \mathcal{C}_G^4,
\end{equation}
(with $\tilde{a}_0=0.237 \pm 0.008$, $\tilde{a}_1=0.674$, and $\tilde{a}_4=4.48$), achieving accuracies of a few percent for a wide range of hadronic EOS. Once $I$ and $M_G$ are measured (e.g., in a binary pulsar), this relation can be used to infer $R$. Although it generally works well, it can become less accurate for EOS exhibiting extreme softening.

Subsequent investigations introduced an alternative dimensionless form of $I$, normalized by $M_G^3$ \cite{Lau:2009bu,Yagi:2013awa,Breu:2016ufb},
\begin{equation}
\frac{I}{M_G^3} = \bar{a}_1 \mathcal{C}_G^{-1} + \bar{a}_2 \mathcal{C}_G^{-2} + \bar{a}_3 \mathcal{C}_G^{-3} + \bar{a}_4 \mathcal{C}_G^{-4},
\label{eq:inertia_hadronic}
\end{equation}
which often yields smaller residuals when compared to $I/(M_G R^2)$. The best-fit coefficients in Ref. \cite{Breu:2016ufb} are 
$\bar{a}_1 = 8.134 \times 10^{-1},$
$\bar{a}_2 = 2.101 \times 10^{-1},$
$\bar{a}_3 = 3.175 \times 10^{-3},$
$\bar{a}_4 = -2.717 \times 10^{-4},$
and indeed capture the behavior of $I$ with slightly higher precision.

We now apply a similar analysis to strange quark stars, described by either the QMDD or the vMIT framework. Initially, we tested whether $I/(M_G R^2)$ provides a universal relation with respect to $\mathcal{C}_G$ in self-bound stars. While the vMIT results largely overlap under this normalization, the QMDD model exhibits noticeable deviations. In contrast, normalizing by $I/M_G^3$ yields a substantially tighter fit; hence, we adopt that normalization in what follows.

\begin{figure}[tb]
\centering
\includegraphics[width=\linewidth]{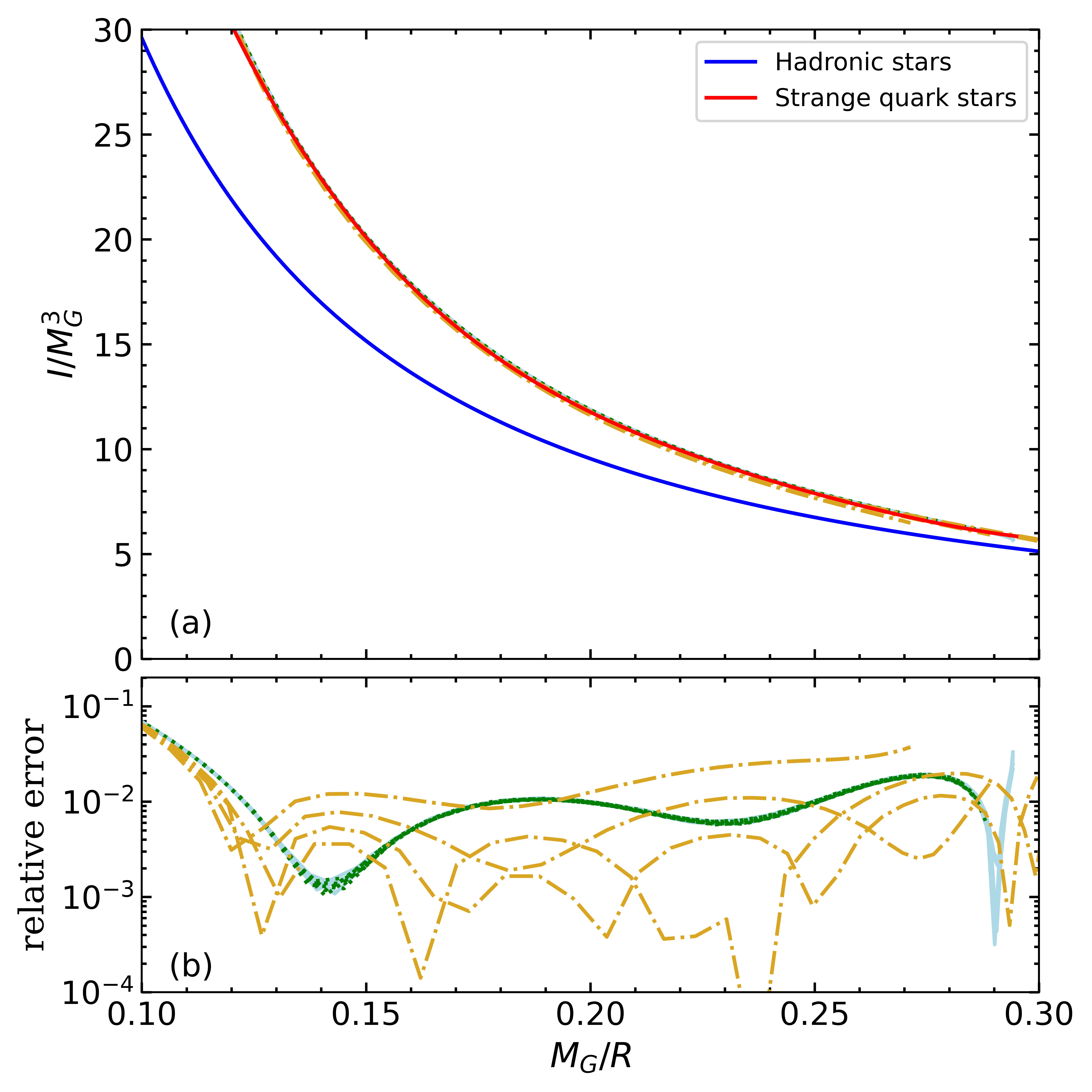}
\caption{(a) Universal relation between the normalized moment of inertia and compactness for slowly rotating strange quark stars. The yellow curves show results from various parameterizations of the vMIT model: $(B^{1/4} [\mathrm{MeV}], G_V  [\mathrm{fm^2}]) = (146, 0.3), (150, 0.2), (154, 0.1), (159, 0)$. The green and light-blue lines represent QMDD model results with $\kappa=0$ and $\kappa=0.5$, respectively. In the QMDD model curves, we fix $a=3$ and vary $C$ so that $e^0_{uds}$ spans 880--930\,MeV in increments of 10\,MeV. The red line is the quartic fit from Eq.~\eqref{eq:inertia_versus_compactness}, and the blue curve represents the hadronic-model fit from Eq.~\eqref{eq:inertia_hadronic}. (b) Relative error of the fit as defined by Eq. \eqref{eq:Error_IoM3}.  }
\label{fig:I_vs_C}
\end{figure}

In Fig.~\ref{fig:I_vs_C}(a), we plot $I/M_G^3$ against the gravitational compactness $\mathcal{C}_G = M_G/R$ for the same parameterizations discussed previously. To quantify the correlation, we fit our numerical data with the quartic expression
\begin{equation}
\frac{I}{M_G^3} = b_4 \mathcal{C}_G^4 + b_3 \mathcal{C}_G^3 + b_2 \mathcal{C}_G^2 + b_1 \mathcal{C}_G + b_0,
\label{eq:inertia_versus_compactness}
\end{equation}
where $b_0 = 172.89$, $b_1 = -2306.9$, $b_2 = 12816$, $b_3 = -33089$, and $b_4 = 32699$. As shown by the red line in Fig.~\ref{fig:I_vs_C}(a), this fit reproduces our results with minimal scatter, improving notably over analogous fits for hadronic stars.

To quantify the robustness of the polynomial fit, we define the relative error as
\begin{equation}
\label{eq:Error_IoM3}
\epsilon_{\mathrm{rel}} \;=\;
\left|
1 \;-\;
\frac{\bigl(I/M_G^3\bigr)_\mathrm{num}}
     {\bigl(I/M_G^3\bigr)_\mathrm{fit}}
\right|
\;.
\end{equation}
Here, \(\bigl(I/M_G^3\bigr)_\mathrm{num}\) is the numerically computed value, while \(\bigl(I/M_G^3\bigr)_\mathrm{fit}\) is the corresponding fit.  In Fig.~\ref{fig:I_vs_C}(b), we show these relative errors as a function of \(M_G/R\).  We find that the fitted polynomial describes the data with high fidelity.

A comparison of the functional forms in Eqs.~\eqref{eq:inertia_hadronic} and \eqref{eq:inertia_versus_compactness} further highlights a notable difference between quark and hadronic fits, reflecting the distinct microphysics of each star type encoded in their respective EOSs. The key distinction is that, in self-bound stars, the pressure vanishes at a very high energy density, so strange quark stars lack the low-density outer layers typical of hadronic stars. This behavior has a pronounced effect on the stellar radius and, in turn, significantly influences the moment of inertia.

To illustrate how these relations can be exploited in practice to discriminate between the two stellar compositions, we adopt the representative values $M_G = 2\,M_\odot$ and $R = 12\ \mathrm{km}$, which correspond to a compactness of $\mathcal{C}_G = 0.25$.  Substituting this compactness into the fits yields
\begin{equation}
\bar I_{\mathrm H}=6.75,
\qquad
\bar I_{\mathrm Q}=7.95,
\label{eq:I_values}
\end{equation}
where $\bar I \equiv I/M_G^{3}$ is the dimensionless moment of inertia.  The two predictions differ by
\begin{equation}
\frac{|\Delta\bar I|}{\bar I}
      \equiv
      \frac{|\bar I_{\mathrm Q}-\bar I_{\mathrm H}|}
           {\tfrac12(\bar I_{\mathrm Q}+\bar I_{\mathrm H})}
      \simeq 16 \% ,
\label{eq:frac_sep}
\end{equation}
providing a quantitative target for future observational efforts aimed at distinguishing hadronic from quark stars through precision measurements of $\bar I$.

The dimensionless moment of inertia can be determined in a relativistic double‐neutron‐star binary through its spin–orbit contribution to the periastron advance. 
PSR~J0737--3039A remains the only known system in which a direct determination of  \(\bar I\) is feasible; although no such measurement has yet been reported \cite{Greif:2020pju}.  Timing projections indicate that after \(\mathcal{O}(10)\) yr of continuous monitoring its moment of inertia can be measured with a conservative uncertainty of \(5\%\text{–}10\%\) \cite{Lattimer:2004nj}.

Because the hadronic and quark predictions differ by \(\sim16\%\) in \(\bar I\) while a single timing measurement is expected to reach only \(5\%\text{–}10\%\) precision, even one high‐quality determination would yield a significance of order $\sim 1.5 - 3~\sigma$, indicating that the two compositions can be distinguished in practice.  For multiple independent measurements, the overall confidence grows as \(\sqrt{N}\), so that just a handful of comparable observations would rapidly elevate the discrimination to \(>5\sigma\) and thereby provide a definitive identification of hadronic versus quark matter in neutron stars.

\subsection{Tidal deformability versus compactness}

\begin{figure}[tb]
\centering
\includegraphics[width=\linewidth]{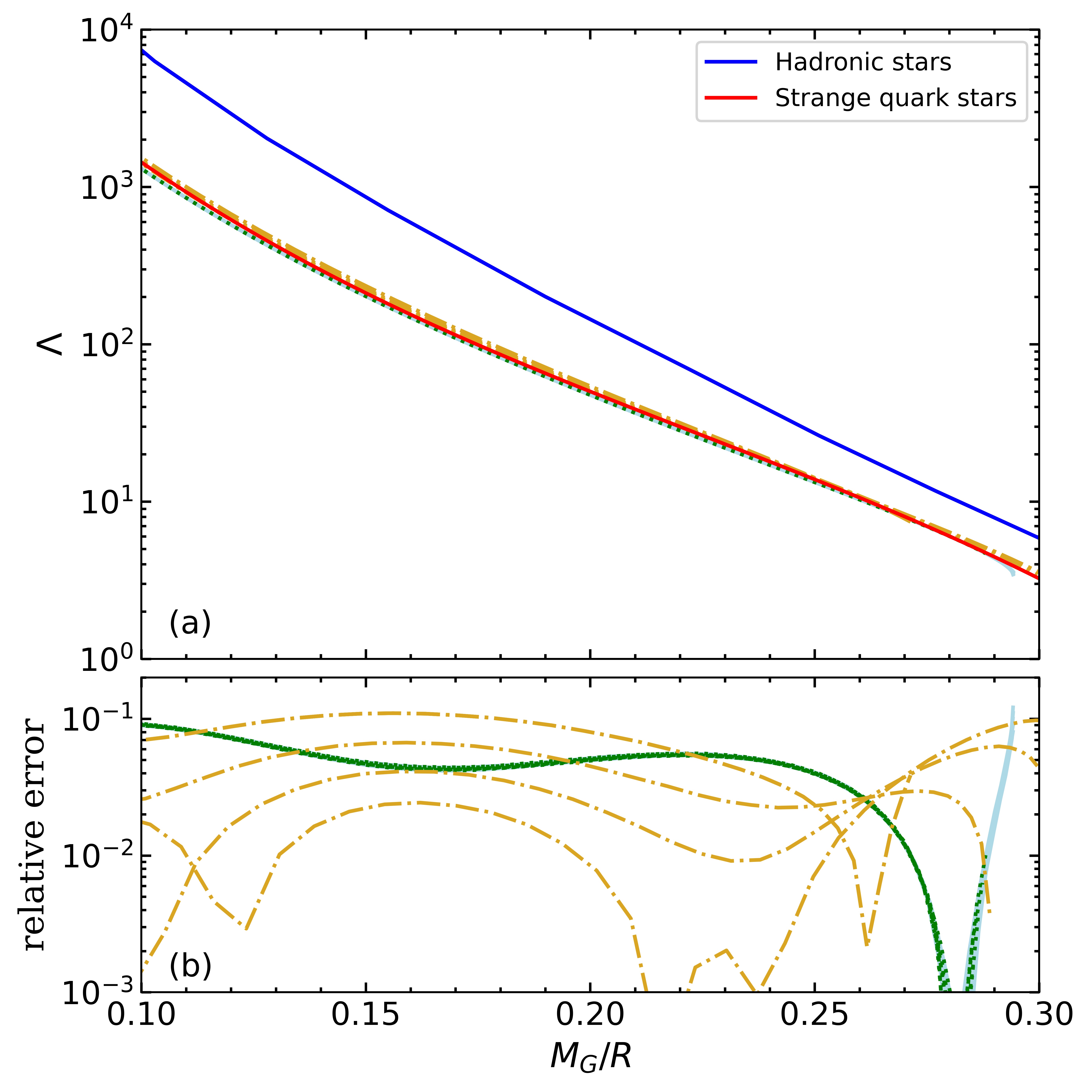}
\caption{(a) Universal relation between tidal deformability and compactness. The solid red line is the polynomial fit given by Eq.~\eqref{eq:fig_LC}, capturing the overall trend across different EOS models and parameter choices. The solid blue line is the hadronic fit presented in Ref. \cite{Li:2023owg}. Other curves follow the color scheme used in the previous figure.  (b) Relative error of the fit as defined by Eq. \eqref{eq:error_fit_Lambda}.  }
\label{fig:tidal_vs_compactness}
\end{figure}

Figure~\ref{fig:tidal_vs_compactness}(a) shows a robust correlation between the tidal deformability $\Lambda$ and the gravitational compactness $\mathcal{C}_G$ that holds for both the vMIT and QMDD models. This universality is notable because it enables one to infer $\Lambda$ from $\mathcal{C}_G$ (and vice versa) without detailed knowledge of the star’s internal structure, making it a valuable tool for constraining compact-star properties via gravitational-wave observations.

To quantify this universal trend, we fit the data in Fig.~\ref{fig:tidal_vs_compactness}(a) with the polynomial
\begin{equation}
\log_{10} \Lambda = a_3 \mathcal{C}_G^3 + a_2 \mathcal{C}_G^2 + a_1 \mathcal{C}_G + a_0,
\label{eq:fig_LC}
\end{equation}
where the best-fit parameters are $a_0 = 5.9886$, $a_1 = -38.861$, $a_2 = 123.93$, and $a_3 = -184.19$. This  fit differs significantly from the existing one for hadronic stars, reflecting the distinct microphysics of self-bound configurations. 

Figure~\ref{fig:tidal_vs_compactness}(b) displays the relative error of our polynomial fit, defined by
\begin{equation}
\epsilon_{\mathrm{rel}} =\left|1 - \frac{\Lambda_\mathrm{num}} {\Lambda_\mathrm{fit}} \right|,
\label{eq:error_fit_Lambda}
\end{equation}
where $\Lambda_\mathrm{num}$ is the numerically computed tidal deformability and $\Lambda_\mathrm{fit}$ is the value from Eq.~\eqref{eq:fig_LC}. Plotting $\epsilon_{\mathrm{rel}}$ against $M_G/R$ shows that the polynomial reproduces the numerical data with high fidelity, indicating that the correlation between $\Lambda$ and $\mathcal{C}_G$ remains robust across different parameterizations of the QMDD and vMIT models.

As a concrete example of how tidal–deformability universality can distinguish between hadronic and quark compositions, consider a compact star with $M_G=2\,M_\odot$ and $R=12\,$km, i.e.\ $\mathcal{C}_G=0.25$.   Substituting this compactness into the fits yields
\begin{equation}
\Lambda_{\mathrm H} = 26.2,
\qquad
\Lambda_{\mathrm Q} = 13.26.
\end{equation}
The two predictions differ by
\begin{equation}
\frac{|\Delta\Lambda|}{\Lambda}
\equiv
\frac{|\Lambda_{\mathrm H}-\Lambda_{\mathrm Q}|}
     {\tfrac12(\Lambda_{\mathrm H}+\Lambda_{\mathrm Q})}
\simeq 65\%.
\end{equation}

The tidal deformability \(\Lambda\) is directly inferred from the phase evolution of binary‐neutron‐star inspirals.  In GW170817,  constraints on the combined parameter \(\tilde\Lambda\) ~\cite{Abbott:2018exr} correspond to an effective uncertainty of order \(50\%\), but next‐generation detectors (Ligo A+, Ligo Voyager, and in particular the Einstein Telescope and Cosmic Explorer) are projected to achieve \(\sim10\%\text{–}20\%\) precision on \(\Lambda\) for individual high‐mass systems \cite{Cho:2022awq}.  Since the hadronic and quark predictions differ by \(\sim65\%\), even a single \(20\%\)‐precision measurement would yield a \(\sim3\sigma\) indication, and multiple independent events would rapidly elevate the discrimination to \(>5\sigma\) confidence.

\subsection{Gravitational compactness versus baryonic compactness}

In this subsection, we examine a new universal relation that links gravitational compactness, $\mathcal{C}_G = M_G/R$, to baryonic compactness, $\mathcal{C}_B = M_B/R$, in strange quark stars. Although universal relations for other macroscopic observables have been extensively studied in the context of hadronic stars, the relationship we find here appears to be novel for quark matter, extending the broader framework of EOS-independent trends.

\begin{figure}[tb]
\centering
\includegraphics[width=\linewidth]{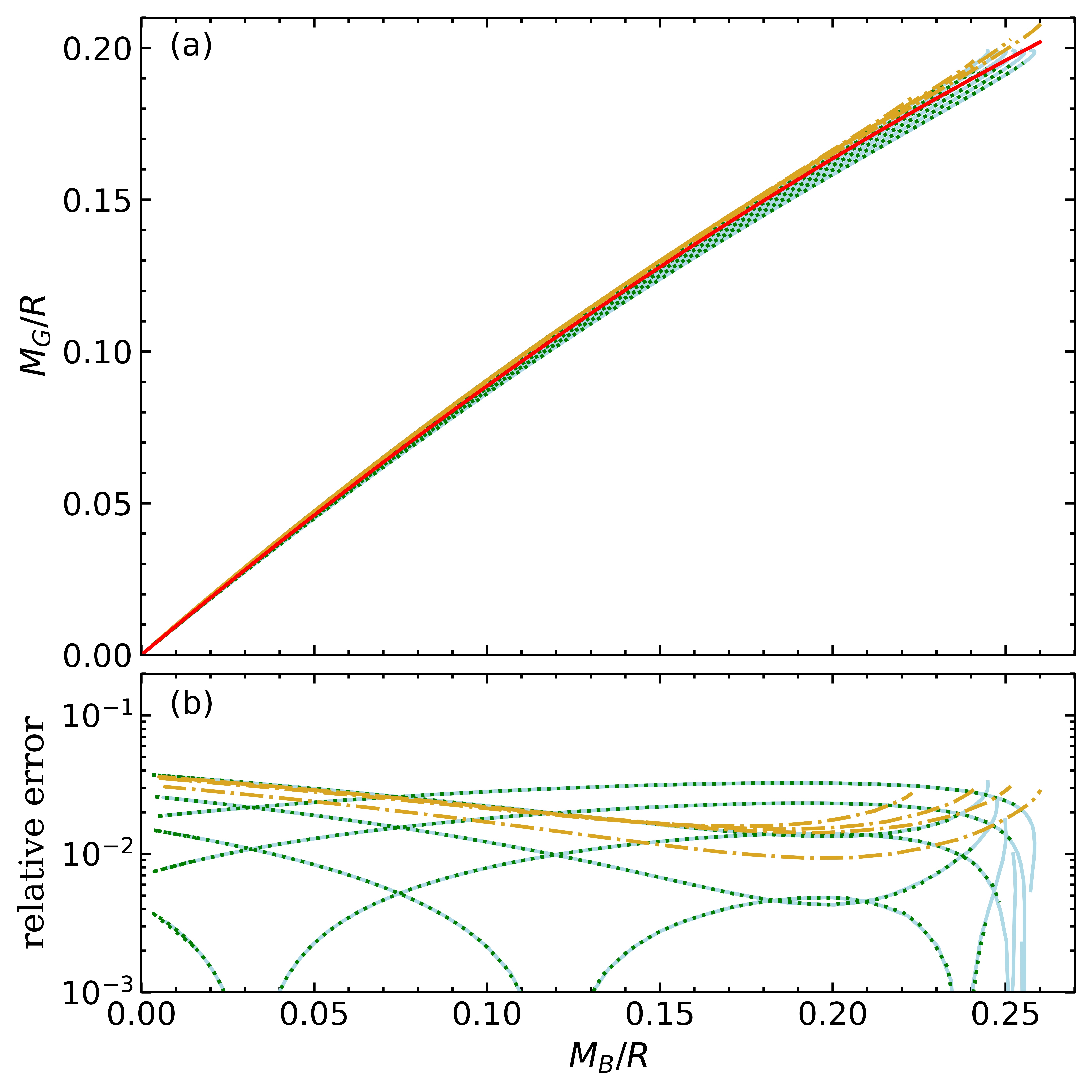}
\caption{(a) Relationship between $M_G/R$ and $M_B/R$, with curves using the same color scheme as in earlier figures. (b) Relative error of the fit, as defined by Eq.~\eqref{eq:error_fit_MB}.}
\label{fig:mass_vs_mass}
\end{figure}

The correlation between $\mathcal{C}_G$ and $\mathcal{C}_B$ follows a quadratic dependence. Figure~\ref{fig:mass_vs_mass}(a) illustrates this behavior for various parameterizations of the QMDD and vMIT models, following the same color scheme as in the previous figures. To describe this relationship quantitatively, we fit the data using
\begin{equation}
\mathcal{C}_G = d_2\,\mathcal{C}_B^2 + d_1\,\mathcal{C}_B,
\label{eq:compactness_relation}
\end{equation}
where $d_2 = -0.682$ and $d_1 = 0.954$. The solid line in Fig.~\ref{fig:mass_vs_mass}(a) shows this quadratic fit, demonstrating excellent agreement with the numerical results.

To assess the accuracy of our fit, we define the relative error by
\begin{equation}
\epsilon_{\mathrm{rel}} = \left|1 - \frac{\mathcal{C}_{G,\mathrm{num}}}{\mathcal{C}_{G,\mathrm{fit}}}\right|,
\label{eq:error_fit_MB}
\end{equation}
where $\mathcal{C}_{G,\mathrm{num}}$ is the gravitational compactness extracted from the numerical data, and $\mathcal{C}_{G,\mathrm{fit}}$ is the corresponding value from Eq.~\eqref{eq:compactness_relation}. Figure~\ref{fig:mass_vs_mass}(b) shows $\epsilon_{\mathrm{rel}}$ as a function of $\mathcal{C}_B$, revealing small deviations over the entire parameter space. This consistency indicates that our quadratic formula accurately captures an EOS-independent relationship between gravitational and baryonic compactness, providing a model-independent means of estimating the binding energy of these objects.

\section{Summary and Conclusions}
\label{sec:conclusions}

In this work, we focused on studying the properties of self-bound SQM and its impact on the structure of strange quark stars. Our main objective was to explore how microscopic aspects of the EOS affect macroscopic observables, while establishing universal and scaling relations that connect these quantities.

We employed two theoretical models to describe the EOS of SQM: the QMDD model with excluded volume corrections and a vector-enhanced MIT bag model. These models incorporate essential features of quark matter in significantly different ways. In the QMDD approach, medium effects enter through an effective quark mass that depends on the baryon number density, providing a self-consistent picture of confinement-like effects and chiral symmetry restoration. Excluded-volume corrections further incorporate repulsive interactions. By contrast, the vMIT model introduces vector repulsion in addition to a bag constant. Through a systematic exploration of the QMDD parameter space ($a$, $C$, and $\kappa$) and the vMIT parameter space ($B$ and $G_V$), we identified in Fig.~\ref{fig:window} the regions where self-bound quark matter is energetically favored over hadronic matter, forming the basis for subsequent analysis.

In Sec.~\ref{sec:stellar_structure}, we summarized standard results used to describe the hydrostatic structure and global properties of compact stars, which laid the groundwork for the universality and scaling studies in the following sections. 

In Sec.~\ref{sec:scaling_relations}, we presented new findings demonstrating how several macroscopic properties of strange quark stars can be directly tied to microscopic EOS parameters. Specifically, we showed that the energy per baryon at zero pressure ($e^0_{uds}$) and the repulsive interaction strength (represented by the excluded-volume fraction $q$ in the QMDD model or the coupling constant $G_V$ in the vMIT model) govern many of these relationships. One key result is that the maximum mass $M_{\mathrm{max}}$ exhibits a linear dependence on $e^0_{uds}$ in the absence of repulsive effects ($G_V=0$ in vMIT or $q=1$ in QMDD). Incorporating repulsive interactions modifies this linear behavior in both models by introducing a common multiplicative factor, as reflected in Eqs.~\eqref{eq:fit_MMAX_QMDD} and \eqref{eq:fit_MMAX_vMIT}.

Section~\ref{sec:universal_relations} highlighted several universal relations among macroscopic stellar properties that remain valid for two substantially different EOS frameworks. These relations—covering the normalized moment of inertia ($I/M_G^3$), tidal deformability ($\Lambda$), and a new correlation between gravitational and baryonic compactness—provide a broader perspective on how global features of strange quark stars can be predicted from minimal knowledge of the EOS. The relative errors across a wide set of model parameters are small, confirming the robustness of the polynomial expressions used to capture these relations.

The relationships derived in this work enable one to extract meaningful constraints on strange quark matter from compact-star observations, such as gravitational-wave data and pulsar timing, without requiring detailed EOS knowledge.
Universal curves in strange quark stars differ markedly from those derived for hadronic stars, owing to the self-bound nature of quark matter versus the confined, nuclear-dominated EOS in hadronic stars.  Consequently, measurements of global stellar parameters, such as tidal deformability, moment of inertia, or baryonic versus gravitational mass, can, in principle, break degeneracies in compact star observations. Such discrimination requires no detailed EOS input and opens a clear path to identifying or ruling out the presence of strange quark matter in astrophysical objects.
Additionally, our scaling relations for the maximum gravitational mass indicate that the mass limit of strange quark stars is largely governed by the overall energy scale at zero pressure, modified by repulsive interactions. Although this idea was implicitly recognized in the earliest studies of strange stars, the novelty here is the explicit inclusion of these parameters in a quite model-independent law. Should future observations place tight constraints on the maximum mass of compact stars, one can invert these relations to constrain both the depth of quark-matter binding and the strength of repulsive effects, thereby offering deeper insight into the fundamental nature of SQM.

\section{Acknowledgements}

GL acknowledges the financial support from the Brazilian agencies CNPq (grant 316844/2021-7) and FAPESP (grant 2022/02341-9). AGG acknowledges the financial support from CONICET under Grant No. PIP 22-24 11220210100150CO,  ANPCyT (Argentina) under Grant PICT20-01847, and the National University of La Plata (Argentina), Project No. X824.

\appendix

\section{The quark-mass density-dependent model}

\textit{EOS for point-like particles.} 
At zero temperature, the energy density, pressure, and chemical potentials of point-like ($\mathrm{pl}$) particles can be expressed in terms of the particle number densities $n_i$ as follows:
\begin{eqnarray}
\epsilon_\mathrm{pl} &=&  \sum_{i} \epsilon_{\mathrm{pl},i} + \epsilon_e 
=  \sum_{i} g M_i^4  \chi(x_i) + \epsilon_e, 
\label{eq:pressure_general} \\ 
p_\mathrm{pl} &=&   \sum_{i} p_{\mathrm{pl},i} + p_e 
= \sum_{i}  \bigl[ g M_i^4 \phi(x_i) - B_i\bigr] + p_e,  \qquad
\label{eq:epsilon_general} \\
\mu_{\mathrm{pl},i} &=& M_i \sqrt{x_i^2 + 1} 
- \frac{1}{3 n_B} \sum_j B_j,   
\label{eq:mu_general} 
\end{eqnarray}
where $B_i$ is the ``bag constant'' defined by
\begin{equation}
- B_i = g \beta(x_i) M_i^3 n_B \frac{\partial M_i}{\partial n_B}.
\label{eq:bag_general}
\end{equation}

The dimensionless Fermi momentum $x_i$ is given by
\begin{equation}
x_i = \frac{1}{M_i} \left(\frac{6 \pi^2 n_i}{g}\right)^{\!1/3},
\end{equation}
where $g$ is the particle degeneracy. The functions $\chi(x)$, $\phi(x)$, and $\beta(x)$ are defined as
\begin{eqnarray}
\chi(x) &=& \frac{x \sqrt{x^2 + 1} \bigl(2 x^2 + 1\bigr) - \sinh^{-1}(x)}{16 \pi^2}, \\
\phi(x) &=& \frac{x \sqrt{x^2 + 1} \bigl(2 x^2 - 3\bigr) + 3 \sinh^{-1}(x)}{48 \pi^2}, \label{eq_definition_of_phi} \\
\beta(x) &=& \frac{x \sqrt{x^2 + 1} - \sinh^{-1}(x)}{4 \pi^2}.
\end{eqnarray}

\textit{Excluded-volume effects.}
Following Refs.~\cite{Lugones:2023zfd, Lugones:2024ryz}, one can incorporate excluded-volume effects into any zero-temperature EOS originally formulated for point-like particles in the Helmholtz representation. We first choose an ansatz for the excluded volume per particle, $b(n_B)$. Based on Ref.~\cite{Lugones:2023zfd}, we define
\begin{equation}
b = \frac{\kappa}{n_B},
\end{equation}
where $\kappa$ is a positive constant. Next, we adapt the expressions for the energy density $\epsilon_{\mathrm{pl}}$, pressure $p_{\mathrm{pl}}$, and chemical potentials $\mu_{\mathrm{pl}, i}$ of point-like particles by rewriting them in terms of the modified variable set $\{ n_j / q \}$ and multiplying by $q = 1 - \kappa$, which is the available volume fraction. The resulting EOS is
\begin{eqnarray}
\epsilon & = &  \sum_{i}  q \epsilon_{\mathrm{pl},i} + \epsilon_e 
= \sum_{i} q g \tilde{M}_i^4  \chi\bigl(\tilde{x}_i\bigr) + \epsilon_e,   
\label{eq:epsilon_with_q}\\
p & = &   \sum_{i}  q p_{\mathrm{pl},i} + p_e 
= \sum_{i}  q [g \tilde{M}_i^4 \phi\bigl(\tilde{x}_i\bigr) - \tilde{B}_i] + p_e,  \qquad 
\label{eq:p_with_q} \\ 
\mu_i & = & \tilde{M}_i \sqrt{\tilde{x}_i^2 + 1}
- \frac{1}{3 \tilde{n}_B} \sum_j \tilde{B}_j, 
\label{eq:mu_with_q}
\end{eqnarray}
where
\begin{eqnarray}
\tilde{n}_B &=& \frac{n_B}{q}, \\
\tilde{M}_i &=& M_i\bigl(\tilde{n}_B\bigr), \\
\tilde{x}_i &=& \frac{1}{\tilde{M}_i} \left[\frac{6 \pi^2 (n_i/q)}{g}\right]^{\!1/3}, \\
-\tilde{B}_i &=& g \beta\bigl(\tilde{x}_i\bigr) \tilde{M}_i^3 \tilde{n}_B \frac{\partial \tilde{M}_i}{\partial \tilde{n}_B}.
\end{eqnarray}

\section{Parameterizing constant-\texorpdfstring{$e^0_{uds}$}{e0uds} lines}
\label{sec:appendix_B}

Curves at fixed values of $e^0_{uds}$, such as the dotted red lines in Fig.~\ref{fig:window}, can be described by simple fits expressed in terms of the EOS parameters. In what follows, we present these fits for both the QMDD and vMIT models.

\subsection{Parameterizations for the QMDD model}

The following expression generates the constant-$e^0_{uds}$ (red) curves shown in Fig.~\ref{fig:window}(a):
\begin{equation}
C =  150\,e^{-3a/5} + 272.8 \left(\frac{e^0_{uds}}{930\,\mathrm{MeV}}\right) - 227.
\end{equation}
The precision of this fit is better than 1\%, and away from $a = 1$ or $a = 4$, its accuracy is noticeably higher.

The energy per baryon at zero pressure as a function of the QMDD model parameters is given by
\begin{equation}
e^0_{uds} = 3.41\,\mathrm{MeV}\,\Bigl[\,C + 227 - 150\,e^{-3a/5}\Bigr].
\end{equation}
Notably, this expression is independent of the parameter $\kappa$ introduced in Eq. \eqref{eq:definition_of_kappa}. Its precision is better than 0.5\%. For consistency, the units of $a$ and $C$ are chosen so that $n_B$ is in $\mathrm{fm}^{-3}$ and $M_i$ is in MeV in Eq.~\eqref{eq:mass_ansatz}.

\subsection{Parameterizations for the vMIT model}

The following expression produces the constant-$e^0_{uds}$ (red) curves shown in Fig.~\ref{fig:window}(b):
\begin{equation}
G_V = \gamma_1 (B^{1/4} - \gamma_2)^{2} - 0.1,
\end{equation}
where
\begin{eqnarray}
\gamma_1 &=& \gamma_{12}\left(\frac{e^0_{uds}}{930\,\mathrm{MeV}}\right)^2 
+ \gamma_{11}\left(\frac{e^0_{uds}}{930\,\mathrm{MeV}}\right) 
+ \gamma_{10}, \quad \\
\gamma_2 &=& \gamma_{21}\left(\frac{e^0_{uds}}{930\,\mathrm{MeV}}\right) 
+ \gamma_{20},
\end{eqnarray}
and
\begin{alignat*}{2}
\gamma_{12} &= 1.1326\times 10^{-2}, \quad & \gamma_{11} &= -2.56525\times 10^{-2},\\
\gamma_{10} &= 1.49756\times 10^{-2}, \quad & \gamma_{21} &= 192.271,\\
\gamma_{20} &= -20.5804.
\end{alignat*}
This parameterization achieves a precision better than 1\%.

The energy per baryon at zero pressure in the vMIT model is then given by
\begin{equation}
e^0_{uds} = \alpha_1\,B^{1/4} + \alpha_2,
\end{equation}
where
\begin{eqnarray}
\alpha_1 &=& \alpha_{12}\,G_V^2 + \alpha_{11}\,G_V + \alpha_{10}, \\
\alpha_2 &=& \alpha_{22}\,G_V^2 + \alpha_{21}\,G_V + \alpha_{20},
\end{eqnarray}
with $B^{1/4}$ and $e^0$ measured in MeV and $G_V$ in $\mathrm{fm}^2$. The coefficients are given by
\begin{align*}
\alpha_{12} &= -2.11,\quad \alpha_{11} = 5.09,\quad \alpha_{10} = 5.60,\\
\alpha_{22} &= 235,\quad  \alpha_{21} = -491,\quad \alpha_{20} = 37.8.
\end{align*}
This final parameterization gives a precision better than 0.2\%.

\bibliography{references} 
\end{document}